\documentclass[11pt,eqsecnum,floats,aps,nofootinbib,prd,tightenlines]{revtex4-2}
\usepackage{hyperref}
\usepackage{amsmath,amssymb,amsfonts,amsthm,amscd}
\usepackage{graphicx}
\usepackage{enumerate}
\usepackage{colordvi}
\usepackage{units}
\usepackage{epsfig}
\usepackage{natbib}
\usepackage{enumerate}
\usepackage{colordvi}
\usepackage{multirow}
\usepackage{afterpage}
\usepackage{subfig}
\usepackage[utf8]{inputenc}

\def\be{\begin{equation}}
\def\ee{\end{equation}}
\def\ba{\begin{eqnarray}}
\def\ea{\end{eqnarray}}

\newcommand{\pv}[1]{\frac{\partial}{\partial #1}}

\def\Lie{\mathfrak{L}}
\def\L{\mathcal{L}}
\def\H{\mathcal{H}}
\def\D{\mathbf{D}}

\def\X{\mathbf{X}}
\def\Y{\mathbf{Y}}
\def\R{\mathbf{R}}
\def\M{\mathcal{M}}

\begin{document}
\title{Scale Symmetry and Friction}
\author{David Sloan}
\email{d.sloan@lancaster.ac.uk}
\affiliation{Department of Physics, Lancaster University, Lancaster UK}
\begin{abstract}
\noindent Dynamical similarities are non-standard symmetries found in a wide range of physical systems that identify solutions related by a change of scale. In this paper we will show through a series of examples how this symmetry extends to the space of couplings, as measured through observations of a system. This can be exploited to focus on observations that can be used distinguish between different theories, and identify those which give rise to identical physical evolutions. These can be reduced into a description which makes no reference to scale. The resultant systems can be derived from Herglotz's principle and generally exhibit friction. Here we will demonstrate this through three example systems: The Kepler problem, the N-body system and Friedmann-Lema\^itre-Robertson-Walker cosmology.

\end{abstract}
\maketitle
\section{Introduction: Poincar\'e's Dream}

In Science and Method \cite{Poincare1914-POISAM}, Poincar\'e invites the reader to consider a world in which the length of all physical objects had been increased a thousandfold, noting that ``What was a meter long would now be a kilometer." Reasoning that an observer could only use the objects that he found within the world as a reference point, he came to the conclusion that the observer would be unable to discern whether such a transformation had taken place at all. It is interesting to note that at the time of Poincar\'e's writing, the length of a meter was defined with respect to a platinum-iridium rod being held at the melting point of ice, and that as such this rod would also be rescaled in his transformation. Thus what previously had the length of a meter would still have this length, as distances would be measured intrinsically. Nonetheless, Poincar\'e's reasoning was that relative to some \textit{absolute} scale the sizes of objects could have changed. This line of reasoning was challenged by, among others, Delboeuf \cite{Delboeuf} who argued that the world we inhabit is not scale invariant, noting that a man whose height measured over a kilometer would lack the strength to be able to walk. A similar counterargument had been previously advocated by Galileo \cite{galilei1954dialogues}, stating that ``...we can demonstrate by geometry that the large machine is not proportionately stronger than the small". 

Despite the counterarguments appearing conclusive, Poincar\'e's dream can be rescued by allowing not only the physical dimensions of the world to be altered, but also the physical constants that determine, for example, the couplings. It is true that on a planet in our universe a thousand times the size of the Earth, with a mass a billion times that of the Earth, a man a thousand times taller than Poincar\'e would not be able to walk, as he would experience a trillion times the force of gravity. With muscles having only a million times the cross-sectional area, he would be unable to overcome this force. However, if the gravitational coupling had one billionth of the strength of that we experience, the ratio of the forces his muscles could exert to that of gravity would be unchanged. It is important to emphasize that this change would constitute not a change of scale within a physical system, but also a change of the physical laws themselves. As such we would not expect any given solution to the theory to be scale invariant, but rather that equivalent descriptions of the same system can be formulated between which the absolute scale of any object would change.

To provide an example of such transformations, let us consider a simple toy model in which two particles in a plane interact subject to a Hooke potential with coupling $k$ (force proportional to separation, $r$) and a Newton potential with coupling $G$ (force inversely proportional to separation squared). As the forces are central, the angular momentum $J=r^2 \dot{\theta}$ is a constant. The equation of motion for the separation of the particles is:
\be \label{HookeNewton} \ddot{r} - \frac{J}{r^3} + \frac{G}{r^2} + k r = 0  \ee
If we are given a solution $(r (t), \theta (t))$ which satisfy this equation for couplings $G, k$. Then we can construct
\be \tilde{r}(t)=\lambda r(\frac{t}{\lambda}) \quad \tilde{\theta}(t)=\theta(\frac{t}{\lambda})\ee
will also satisfy the equation for couplings $\tilde{G} = \lambda G$ and $\tilde{k} = \frac{k}{\lambda^2}$. Thus an observer who plotted the shape orbit traced out by the particle, would be unable to detect any change if $r,G$ and $k$ were changed in this manner. One might object at this point to the rescaling in time - one system would evolve more slowly in $t$ than the other. However, this time must also be read from a physical device; if we equip our observer with a pendulum clock, for example, with time period $T=2\pi \sqrt{\frac{l}{g}}$, to follow Poincar\'e we must set $\tilde{l} = \lambda l$, for all lengths to be affected equally, and $g$ derives from the radius of the Earth, its mass and the Newton potential. Rescaling all of these (keeping the density of the Earth fixed) we see that the clock would now have time period $\tilde{T} = \lambda T$. Thus the angular distance covered in one tick of the pendulum clock would be unchanged.

The example above is quite simple to generalize to include more general couplings. We can express a general potential in the form 
\be V(r,\theta) = \sum_i C_i r^i V_i(\theta) \ee
in which case we find an analogous rescaling holds if allow the couplings $C_i$ to be rescaled to $\tilde{C}_i=\lambda^{-i} C_i$. Furthermore, it is a straightforward exercise to generalize to the case where we let $\tilde{r}=\lambda r(\frac{t}{\lambda^n}), \tilde{\theta}= \theta(\frac{t}{\lambda^n})$, and those where we introduce more particles. 

A natural question to ask at this point is that given that such transformations are undetectable to an observer, can we formulate our models to work only in terms of entities which are invariant? Such a move is at the heart of shape dynamics \cite{Shapes1,Shapes2,Shapes3,Shapes4}, which aims to describe physical phenomena without ever referring to scale. Surprisingly, it turns out that this is indeed possible. In doing so we replace the usual symplectic geometry of Hamiltonian systems with `contact geometry' \cite{ContactIntro,Bravetti,Leon,Bravetti2,DynSim}. Further, we will see that we can derive the dynamics of these observables from an action principle that is expressed only in terms of the invariants themselves. 

This general set of symmetries are referred to as \textit{Dynamical Similarities} \cite{DynSim,Bravetti:2020tau,Ellis:2019coe}. It has been shown that such symmetries are commonplace in physical systems and have particularly important philosophical and mathematical implications in the cosmological sector \cite{Gryb:2020wat,Gryb:2021qix,Sloan:2019wrz,Sloan:2020taf}.  In previous work we have dealt with the case where a single force was acting, and chosen the appropriate symmetry that kept the coupling of the relevant potential fixed. In this paper we will lay out the general case in which multiple forces are acting on any number of particles, and show how this further applies to homogeneous, isotropic cosmological systems with multiple matter sources present. 

Throughout this work we will use superscripts to denote powers of objects, and indexing labels will all be subscripts, regardless of the mathematical object in question. This goes somewhat against the usual convention of denoting elements of the tangent bundle or cotangent bundle differently with upper of lower indices, but will help disambiguate between, for example, the n-th momentum element of a set, $p_n$ and the n-th power of momentum, $p^n$. 

\section{Lagrangian and Herglotz Mechanics}
\label{Sect:LagMech}
Let us recapitulate the usual scheme of Lagrangian mechanics\footnote{In this paper we will restrict ourselves to the case of Lagrangians which are time independent, and depend only on positions and their first derivatives. This can all be generalized to include time dependence and higher derivatives, but to do so would introduce unnecessary clutter to our presentation.}, and contrast with the generalizations provided by Herglotz. In this section we will only present a brief account of the standard results, full proofs of which can be found in \cite{geiges2008introduction,Carinena:2013zpa,Carinena:2014bda,ArnoldBook}. We consider a physical system defined on the tangent bundle over a configuration space, $M=TC$ which typically is written in terms of positions $q_i$ and their velocities, $\dot{q}_i$. Physical trajectories are those which extremize an action across a set of curves through the space of $q_i, \dot{q}_i$, where the action is the integral over time of a Lagrangian, $\L$, which is a function of these positions and velocities. 
\be S = \int_{t_i}^{t_f} \L(q,\dot{q}) dt \ee
Extremization means that $\delta S=0$, and after a textbook calculation we arrive at the usual Euler-Lagrange equations
\be \label{EL} \frac{d}{dt} \left(\frac{\partial \L}{\partial \dot{q}_i} \right) - \frac{\partial \L}{\partial q_i} = 0 \ee
Given such a Lagrangian, we can construct the Hamiltonian $\H$, a function on phase space, the cotangent bundle $M'=T^*C$, through a Legendre transform. Supposing $\frac{\partial^2 \L}{\partial \dot{q}_i \partial \dot{q_j}}$ is positive definite, then we can uniquely set $p_i = \frac{\partial \L}{\partial q_i}$ and render $\H$:
\be \label{Hdef} \H = \sum_i p_i \dot{q}_i - \L \ee
from which we find Hamilton's equations: $\dot{q}_i=\frac{\partial \H}{\partial p_i}$ and $\dot{p}_i =-\frac{\partial \H}{\partial q_i}$ and a straightforward calculation shows that $\H$ is a constant:
\be \label{Hdot} \frac{d\H}{dt} = \sum_i \left( \frac{\partial \H}{\partial q_i}\dot{q}_i + \frac{\partial \H}{\partial p_i} \dot{p}_i \right) = \sum_i \left(\dot{p}_i\dot{q}_i - \dot{q}_i\dot{p}_i  \right) = 0 \ee
In equation (\ref{Hdef}) we introduced the symplectic potential, $\theta=\sum_i p_i dq_i$, the exterior derivative of which is the symplectic form.  On phase space the symplectic form, $\omega=\sum_i dp_i \wedge dq_i$ can be used to form a `natural' measure, $\Omega=\omega^{\wedge n}$ where $n$ is the dimension of the configuration space \cite{ArnoldBook,Measure,Measure2}. This is the basis of much of statistical mechanics in the Hamiltonian formalism. A key result is Liouville's theorem, that $\omega$ is constant in time. The phase space volume occupied by a set of solution as measured by $\Omega$ thus does not change as the system evolves. Thus time independent Hamiltonians (and their associated actions) describe conservative systems.

Let us now consider the generalization of these systems to include the action, $S$ as part of the Lagrangian $\L^H(q,\dot{q},S)$, hence $\L^H$ is a function on $TM\times \mathbb{R}$. These we first considered by Herglotz  as a way to introduce non-conservative terms, and have found a variety of applications \cite{deleon2021herglotz,SMJCMdL01,MRUGALA1991109,Garra_2017,Lazo_2018,P_S_Santos_2018} across applied mathematics. Hence, we shall call these `Herglotz Lagrangians'. Extremizing the action we find equations of motion similar to the standard Euler-Lagrange equations (\ref{EL}) and reduce to them in the case where $\L^H$ is independent of $S$:
\be \label{HL} \frac{d}{dt} \left(\frac{\partial \L^H}{\partial \dot{q}_i} \right) - \frac{\partial \L^H}{\partial q_i} - \frac{\partial \L^H}{\partial S}\frac{\partial \L^H}{\partial \dot{q}_i} = 0 \ee
The important distinction here is the extra `frictional' effect introduced by the $\frac{\partial \L^H}{\partial S}$ term. We refer to this as being frictional as it is the origin of the non-conservative terms; Consider the Hamiltonian formed following a Legendre transformation per equation (\ref{Hdef}), again introducing $p_i=\frac{\partial \L^H}{\partial \dot{q}_i}$. This is a function on the contact manifold, $T^*M \times \mathbb{R}$, an odd dimensional space, and we shall refer to such Hamiltonians as `contact Hamiltonians'. Given a contact Hamiltonian, $\H^c$, the equations of motion are:
\be \label{CHEoM} \dot{q_i} = \frac{\partial \H^c}{\partial p_i} \quad \dot{p_i} = -\frac{\partial \H^c}{\partial q_i} - p_i \frac{\partial \H^c}{\partial S} \quad \dot{S} = p_i \frac{\partial \H^c}{\partial p_i} -\H^c \ee
From these equations the frictional nature of such systems becomes more readily apparent; a direct calculation following equation (\ref{Hdot}) reveals:
\be \label{dissipation} \frac{d\H^c}{dt} = \sum_i \left( \frac{\partial \H^c}{\partial q_i} \frac{\partial \H^c}{\partial p_i}  - \frac{\partial \H^c}{\partial p_i} 
\frac{\partial \H^c}{\partial q_i} - p_i \frac{\partial \H^c}{\partial S}\frac{\partial \H^c}{\partial p_i} + p_i \frac{\partial \H^c}{\partial S} \frac{\partial \H^c}{\partial p_i} - \H^c \frac{\partial \H^c}{\partial S} \right)  = -\H^c \frac{\partial \H^c}{\partial S} \ee
The counterpart to the symplectic potential on a contact manifold is the contact form, $\eta=-dS + \sum_i p_i dq_i$. Since the contact manifold is odd dimensional, we cannot form a measure on it simply by using powers of the equivalent of symplectic form, $d\eta$. However, $\Theta=\eta \wedge d\eta^{\wedge n}$ is a volume form on contact space. We can see the non-conservative nature of contact Hamiltonians and their Herglotz descriptions from the evolution of $\eta$; 
\be \label{MeasureDissipation} \dot{\eta} = -\frac{\partial \H^c}{\partial S} \eta \rightarrow \dot{\Theta} = -(n+1) \frac{\partial \H^c}{\partial S} \Theta \ee
This means that the volume on the contact manifold occupied by a set of solutions is not conserved through evolution, but rather undergoes focussing/spreading. This allows us to understand the apparent contradictions found in measures on spaces of inflationary cosmological solutions, as the observables form a contact manifold - see \cite{Corichi:2010zp,Corichi:2013kua,Measure,Measure2} for details. 

To illustrate the results above, let us consider the description of a damped harmonic oscillator. The Herglotz-Lagrangian is similar to the Lagrangian of a simple harmonic oscillator, but with a term linearly proportional to the action added:
\be \label{LDHO} \L^h  = \frac{m \dot{x}^2}{2} - \frac{kx^2}{2} - \frac{\mu S}{m} \ee
from which the Herglotz-Lagrange equation (\ref{HL}) is:
\be \label{DHOEOM} m\ddot{x} + \mu \dot{x} + kx = 0 \ee
which is the equation of motion of a damped harmonic oscillator. The extra term, $\mu \dot{x}$ arises as a result of the action term in $\L^h$, and gives rise to energy dissipation in the evolution of the system. The contact Hamiltonian, $\H^c$, follows from the Legendre transform, with $p=m \dot{x}$:
\be \H^c = \frac{p^2}{2m} + \frac{kx^2}{2} + \frac{\mu S}{m} \ee
which has equations of motion 
\be \dot{x}=\frac{p}{m} \quad \dot{p} = -kx -\frac{\mu p}{m} \quad \dot{S}=\frac{p^2}{2m} - \frac{kx^2}{2} - \frac{\mu S}{m} \ee
equivalent to equation (\ref{DHOEOM}). Either by direct calculation from the equations of motion, or from equation (\ref{dissipation}) we can see that the rate of energy loss is $\frac{d\H^c}{dt} = - \frac{\mu \H^c}{m}$. This is unsurprising as solutions to the equations of motion show that the system asymptotes towards a stationary point at the origin. Physically we see that the frictional terms remove mechanical energy from the system, leaving the oscillator to tend towards resting at the minimum of its potential, $x=0$. 

To highlight the dissipative nature of the system let us examine the evolution of the volume form $\Theta = -dS \wedge dp \wedge dx$. Again from direct calculation (an informative exercise for the reader) or from equation \ref{MeasureDissipation} we see that:
\be \dot{\Theta} = - \frac{2\mu}{m} \Theta \ee
and hence the volume of the contact space occupied by a set of solutions will reduce over time. On the space of solutions this can be understood as all solutions asymptote towards the same endpoint, hence the volume they occupy on the contact space should contract to that point. From a physical perspective  this is simply the friction leading all such oscillators to the same resting point. 

\section{The Kepler Problem}
\label{Sect:KeplerExample}

Let us consider a dynamical system consisting of a two body system with a Newtonian central force. This we will treat as a toy model in which we demonstrate the mathematical tools which allow for our analysis, and present the logic behind our arguments. Our ultimate goal is to express the behaviour of systems purely in terms of measurable quantities. A step in this is to demonstrate the representational redundancies in expressing the behaviour of a bound pair of particles as measured by an observer using a second such (approximately isolated) pair as a rod and clock. Initially, however, we will avoid the associated mathematical clutter by a single two body system in reduced variables, and show how the system can be expressed in these terms.

The solutions to the two body system are known to be conic sections - circles, ellipses, parabolas and hyperbolas. We will focus on bound systems - ellipses and circles, as hyperbolas and parabolas will not work well as rods and clocks. Since we will be interested in using a pair of particles as a measuring apparatus, we will posit that the separation of the particles, $r$, is not directly measurable. The angle $\theta$ will be treated as directly observable, as it may be thought of as being measured against some fixed background of stars, which for our purposes are taken to be at an unmeasurable distance - the role of the stars is simply to give us access to measuring the angle $\theta$. Therefore to a complete solution is expressed in terms of the eccentricity of the orbit, the angle of the major axis. Further, in comparing two such ellipses the phase of the orbit would be observable. Thus the space of such solutions is three dimensional: $S^1 \times S^1 \times [0,1)$. \footnote{Note that the interval can be replaced by $\mathbb{R}_+$ by working in $q=\frac{e}{1-e}$ for example; it is only really the topology of the space of solutions that is important.} 

For simplicity we will set the masses of the particles to unity, and work in the center of mass frame. The Lagrangian for such a setup is:
\be \label{KeplerLagrangian} \L = \frac{\dot{r}^2}{2} + \frac{r^2 \dot{\theta}^2}{2} + \frac{C}{r} \ee
This gives rise to two second-order differential equations in $r$ and $\theta$. Thus in order to determine a solution, one must specify five pieces of information; the values of $r,\dot{r},\theta,\dot{\theta}$ and $C$. This over-describes the space of solutions. The question then arises as to whether these redundancies can be removed at the level of the dynamical system. Can we describe the system purely in terms of quantities that describe distinct solutions?

The dynamics is derived from an action principle, the minimization of which provides the equations of motion. We must further specify the value of the constant $C$, and the initial conditions.  As we have noted previously, however, under transformations $D$ in which $D:S \rightarrow \lambda S$, the equations of motion for the invariants of $D$ are unchanged, as minimization of the action is unaffected by the transformation. In other words, if $\delta S=0$ then $\delta DS = \lambda \delta S = 0$. Hence a transformation which leaves the observables unchanged, but under which $\L\rightarrow \lambda \L$ has indistinguishable solutions. In this case we will want $D$ to leave angles and the eccentricity unchanged.

In previous work we have considered the case where $D$ acts only on the tangent bundle of the configuration space through altering $r$ and $\dot{r}$ (or in the Hamiltonian case, the cotangent bundle through scaling $r$ and $P_r$). Here we will extend this analysis to also allow transformations that alter the interaction strength $C$. Our motivation is that ultimately all such interactions must be measured through experiment. As such, the value of $C$ is to be determined by observations of $r,\theta$ (and possibly $t$, which in turn should be a function of the observables). Hence a sympathetic transformation which alters not only $r$ and $t$ but also $C$ in such a way that the relational motion is unaffected should not be apparent to an observer. 

If we consider a direct scaling, $D: (\theta,r,t,C) \rightarrow (\theta,\alpha r, \beta t, \gamma C)$ we see that the requirement that we rescale the Lagrangian reduces to $\alpha^2 = \beta^2 \gamma$. The case in which we kept $C$ constant is apparent when we fix $\gamma=1$, and the invariance of the system under time reversal is reflected by the fact that $\beta$ appears squared; choosing $\pm \beta$ is indistinguishable. We can parametrize most of the space of transformations purely in terms of $\alpha$ which sets $\beta=\alpha^\chi$ and $\gamma=\alpha^{2-2\chi}$, for any real number $\chi$. There are three special choices. $\chi=0$ fixes the rate of time of the system, but alters the size and the coupling $C$. $\chi=1$ fixes the coupling $C$ and rescales both size and time. Finally, a third choice is to fix $\alpha=1$, in which case the transformations are given $\gamma=\beta^{-2}$, which keeps size fixed and alters the time and coupling. We note here that since we leave the angle $\theta$ invariant, both the kinetic terms have the same scaling. 

We can equivalently express our systems in terms of a Hamiltonian, $\H$ which is a function on the cotangent bundle, and the symplectic form $\omega$. Following a Legendre transformation, these are given
\be \label{SymplecticKepler} 0:=\H = \frac{P_r^2}{2} + \frac{P_\theta^2}{2r^2} - \frac{C}{r} - E \quad \omega =dP_r \wedge dr + dP_\theta \wedge d\theta \ee
Since the Hamiltonian is a constant in this set-up, we have introduced the total energy of the system, $E$, and hence the system evolves on the constraint surface $\H=0$. The space of possible solutions to this system $\M$ is the product of the cotangent bundle $T^*Q$ with two copies of the positive reals (one each for $E$ and $C$), subject to the constraint. Choosing initial conditions and specifying a value of $C$ uniquely determines $E$. We can characterise the transformations $D$ in terms of a vector field $\D$ on $\M$ which leaves the constraint surface unchanged. In other words, the action of $D$ on the $r,t$ and $C$ is equivalent to the action of $\D$ on $\M$ as both map between equivalent descriptions of the same solution. To retain the dynamics, these transformations can only act on the symplectic form by linear rescaling. For simplicity we normalize the length of $\D$ such that $\Lie_\D \omega = \omega$. We will label such transformations in terms of the scaling effect they have upon the length $r$:
\be \D_\zeta:= P_\theta \frac{\partial}{\partial P_\theta} +\zeta r\frac{\partial}{\partial r} + (1-\zeta) P_r \frac{\partial}{\partial P_r} + (2-\zeta) C \frac{\partial}{\partial C}+ (2-2\zeta) E \frac{\partial}{\partial E} \ee
We see $\Lie_{\D_\zeta} \H = (2-2\zeta)\H$. Note that by construction $\D_\zeta$ leaves invariant angles $\theta$ and it is easy to verify the eccentricity is also unchanged ($\Lie_{\D_\zeta} e = 0$). We again note that there are special choices that can be made here: $\zeta=0$ leaves fixed the length scale $r$, $\zeta=1$ fixes the energy, and $\zeta=2$ leaves fixed the coupling. Using these fields we can move along the constraint surface without altering the observable quantities. This can be useful in translating between conventions; we could choose our solutions to be those on which $C=1$ and $E=-1$ for example, and thus we can take any solution and Lie-drag it along $\D_1$ until $C=1$ and then along $\D_2$ until $E=-1$. We start with a six-dimensional space $\M$. Imposing the constraint $\H=0$ reduces this by one, and $\D_1$ and $\D_2$ move in a two dimensional plane within this surface. Thus we are again left with a three-dimensional space of distinguishable solutions.

Let us now turn our attention to describing dynamics in terms of these distinguishable variables. Since moving along any of the $\D_\zeta$ does not change the observables, we will work with the set of invariants of one of these; for the sake of simplicty we pick $\D_1$. We will first use $\D_2$ to fix $E=-1$, and we work with variables $x$ for which $\Lie_{\D_1} x = 0$. These form an algebra with basis $A=P_r$, $\theta$, $B=\frac{P_\theta}{r}$, and $\mu=\frac{C}{r}$. It was shown in \cite{DynSim} that the dynamics of a Hamiltonian system with such a symmetry is equivalent to that derived from a contact Hamiltonian. In this case the contact Hamiltonian and contact structure are:
\be \label{ContactKepler} 0=:\H^c = \frac{A^2}{2} + \frac{B^2}{2} -\mu +1 \quad \eta = -dA +Bd\theta + \mu dz \ee
From this and the definition in equation (\ref{Reebdef}) we see that the Reeb vector field is $R=\frac{\partial}{\partial A}$. Since $\mu$ is composed of both a constant piece ($C$) and a piece that varies in time ($r$), it is dynamical. In our system this dynamical evolution is accounted for by promoting $\mu$ to being a momentum on an extended contact manifold. In other words, we have taken $\mu$ and considered it as the momentum conjugate to a dummy configuration variable, $z$. We do this illustrate the equations of motion more simply. As we shall see below, $\mu$ actually plays a role that is similar to the constant energy in a symplectic system. As we show in equation (\ref{dissipation}), had we set the constraint $\H^c$ to a non-zero value, this value would evolve in time, and the equation of motion for this is precisely that of $\mu$. The equations of motion can be found from equation (\ref{CHEoM})
\be \theta '=B \quad B'=-AB \quad \mu '=-A\mu \quad A'=\frac{B^2}{2}-\frac{A^2}{2}+1 \ee
Under the identification $x'=r \dot{x}$, we see that this system encapsulates the same dynamics as that described by equations \ref{SymplecticKepler}. However for a complete solution here we need only specify three pieces of information at a given event: $\theta, A,B$ together with the constraint uniquely determine $\mu$ at this point, and thus can be used to integrate the system. Let us now explicitly solve this system. First we note that the equations of motion for $B$ and $\mu$ allow us to note that:
\be \frac{dB}{d\mu} = \frac{B'}{\mu'} = \frac{B}{\mu} \rightarrow B = \lambda \mu \ee
for some constant $\lambda$. Then we can express the motion in terms of $\theta$, to see:
\be \frac{d^2 B}{d\theta^2} = \lambda - B \quad A=-\frac{dB}{d\theta} \ee
and hence 
\be B= B_0 \cos(\theta-\theta_0) + \lambda \quad A = -B_0 \sin(\theta-\theta_0) \ee
and the Hamiltonian constraint gives:
\be B_0^2 = \lambda^2 -2 \ee
and we arrive at a description of the system in terms of the eccentricity, $e=\frac{B_0}{\sqrt{B_0^2+2}}$ and the angle to perihelion, $\theta_0$. It is a simple exercise to show that had we chosen to take the total energy of the system as described in equation \ref{SymplecticKepler} to be negative we would have found the equivalent result for parabolic/hyperbolic trajectories. 

Let us briefly recapitulate the information that we have required to solve our system. To solve the system in the usual Hamiltonian (or Lagrangian) formulation we would have to specify numerical values for the radial momentum, the angular momentum, an initial radial displacement,an initial angle and the strength of the Newtonian coupling. Upon solving the system we would see that the observable quantities under-determine this data; we would need further input from measuring apparatus (a rod and a clock, for instance) to determine these. We could have judiciously chosen to set our initial conditions at a specific event (for example, the vanishing of radial momentum at aphelion) but we would still require external inputs to solve our system. Thus by initially eliminating the redundancy under dynamical similarity we arrive at a more parsimonious description of physics, which crucially only relies upon distinguishable, observable quantities. This system is autonomous; it needs no further external inputs to be fully integrated. Once integrated we can choose to introduce a scale by giving the separation of the particles some value at a given event, and solve for the evolution of this scale by quadrature to recover the usual description of dynamics. However we reiterate that this is not necessary to find the complete set of observables. 

We can recover the original Hamiltonian symplectic system from the contact system through symplectification \cite{ArnoldBook}. Since the contact form is defined up to a multiplicative factor which we will, with foresight label $R$, we can promote this factor to a coordinate on configuration space, and arrive at a symplectic form through
\be \omega_s = d(R\eta) = dA \wedge dR + d(RB) \wedge d\theta + d(R\mu) \wedge dz \ee
and taking this to be a canonical symplectic form following $d\Pi \wedge dq$ for momenta $\Pi$ and coordinates $q$, this sets
\be \Pi_R = A \quad \Pi_\theta = RB \quad \Pi_z = R\mu \ee
Rendered in these variables, the contact Hamiltonian (from equation \ref{ContactKepler}) is:
\be H^c = \frac{\Pi_R^2}{2} + \frac{\Pi_\theta^2}{2 R^2} - \frac{\Pi_z}{R} - 1 \ee
which in turn is the usual symplectic system given in equation \ref{SymplecticKepler}, with the constant $C$ promoted to a momentum which the equations of motion leave constant. Thus we see that the usual symplectic system is just the symplectification of the underlying contact system. For this reason the symplectic system must, necessarily be under-determined. The observable quantities were all described by the contact system, and in symplectification we introduce a choice of the scale given by $R$. 

As advertised, the contact system described in equation \ref{ContactKepler} has a Lagrangian equivalent described by following Herglotz' principle. The dynamics can be derived from an action principle which can be stated as ``Extremize $A$ subject to $A'=\frac{\theta'^2}{2}-\frac{A^2}{2}+1$ given initial conditions for $\theta, \theta'$ and $A$." Since $A=\int A' d\tau$, the equation of motion for $A'$ plays the role of a Lagrangian. We show in appendix \ref{Appendix} how the equations of motion for $\theta$ can be explicitly derived from this principle. Herein we see that the complete system is describe only in terms of a three dimensional set. From extremizing $A$ we find the equation of motion for $\theta: \theta'' = -A\theta'$. Treating $A'$ as a Lagrangian, we find that the associated Hamiltonian is exactly that given by equation ($\ref{ContactKepler}$).

\section{Generalization of the Kepler Problem}
\label{Sect:KeplerGen}

Let us now consider a more general scenario than that of the Kepler system above. We will allow for potentials that have general power laws in $r$ and also angular dependence. The more general form of potentials allows us to prepare for multiple interacting bodies since the forces acting on a particle will depend not only on its separation from other particles, but also the relative locations of those particles. Including different power laws in $r$ addresses the case of multiple forces acting. Likewise this encompasses models in which we approximate a complex object (such as an atom) as a particle and expand the potential experienced by another particle in terms of a multipole expansion. 

We consider a Lagrangian of the form:
\be \label{LagrangianGeneral} \L = \frac{\dot{r}^2}{2} + \frac{r^2 \dot{\theta}^2}{2} - \sum_{i:V_i \neq 0} C_i V_i(\theta) r^i \ee
in which the sum is taken over $i$ such that the corresponding $V_i$ is non-zero. Hereafter we shall drop the angular dependence in $V_i$. It is displayed above so that the explicit form of the general potential can be seen. The $C_i$ are coupling coefficients. This does introduce a degeneracy into our description as rescaling both the angular component of the potential and its associated coupling in reciprocal would not change the functional form of the Lagrangian. To fix this we are free to impose any normalization condition on the angular part of the potential, such as setting the coefficient of the lowest non-zero term in the Fourier series to unity. As an example, we could set $V_{-2}=\cos(\theta)$ to model the effects of a dipole moment.

The Hamiltonian of this system is:
\be \label{HamiltonianGeneral} \H = \frac{P_r^2}{2} + \frac{P_\theta^2}{2r^2} + \sum_{i:V_i \neq 0} C_i V_i r^i \ee
and we can combine the total energy of the system into $C_0$ with appropriate choice of a constant term in $V_0$. As noted in the Kepler example, we consider $\H$ to live on a the product of the phase-space of the system with one copy of the positive reals for each non-zero coupling constant; $\M=T^*Q \times \mathbb{R}_+^k$. There is an important distinction to be made as although we will see that transformations $D$ can move between systems with differing coupling constants, these transformations respect sign of each coupling. This should be apparent in the Kepler case, as the solutions in which the coupling is attractive, repulsive and simply not present, have physically (topologically) distinct solutions. As an example, closed circular orbits are not solutions to the Kepler problem if the coupling is either repulsive or not present at all. 

We can again parametrize transformations in terms of their scaling effect upon length. These are thus given:
\be \label{GenKeplerDS} \D_\zeta:= P_\theta \frac{\partial}{\partial P_\theta} +\zeta r\frac{\partial}{\partial r} + (1-\zeta) P_r \frac{\partial}{\partial P_r} + \sum_{i:V_i \neq 0} (2-(2+i)\zeta) C_i \frac{\partial}{\partial C_i}\ee
For simplicity we will again choose to work with the invariants of $\D_1$. This is a choice we make for reasons of simplicity of the mathematical representation. Other choices may be more useful in certain physical situations. In the case of a total collision, for example, the highest negative power of $r$ in the potential would be expected to dominate dynamics, and hence it would be appropriate to work in a system in which its coupling was kept fixed. 

As in the Kepler example, in the contact system the couplings become dynamical objects. We let $\mu_i=C_i^{-\frac{1}{i}}r^{-1}$ for $i \neq 0$, and $\mu_0=C_0$, and can thus render the contact system in terms of its Hamiltonian and contact form:
\be \label{ContactGeneral} 0=:\H^c = \frac{A^2}{2} + \frac{B^2}{2} + \mu_0 + \sum_{i:V_i \neq 0} \mu_i^{-i} V_i \quad \eta = -dA +Bd\theta + \sum_i \mu_i dz_i \ee
Once again we have introduced a set of dummy configuration variables $z_i$, and written $A=P_r$ and $B=\frac{P_\theta}{r}$. The equations of motion for this system are simple when expressed in terms of these variables. Since each of our coupling encoding momenta are conjugate only to dummy variables, their evolution can expressed in terms of themselves and the effective frictional term $A$ alone.
\be \label{GenMuEom} \mu_i' = -A \mu_i \rightarrow \log(\mu_i) ' = -A \ee
%
%
\be \label{GenBEom} B'=-AB - \sum_{i:V_i \neq 0} \mu_i^{-i} \frac{\partial V_i}{\partial \theta} \ee
Here we reiterate that although one would not expect explicit $\theta$ dependence in the two body problem, we have retained the general form of these terms to provide insight into the more general $n$-body problem in which angular distributions will matter. 

Finally the equation of motion for $A$ can be found:
\be \label{ContactGeneralA} A'=\frac{B^2}{2}-\frac{A^2}{2} - \sum_{i:V_i \neq 0} (1+i)V_i \mu_i^i \ee
Hence we have a complete system in which there is a closed dynamical system described by the three prior variables, $A$,$B$ and $\theta$ alongside the $\mu_i$ which encapsulate the information about the couplings of the system. Again this requires one fewer initial data points than the usual Hamiltonian representation to close. In that case we would need the values of each of the $C_i$ alongside three of $r,P_r,\theta,P_\theta$, with the fourth found through the Hamiltonian constraint. In the contact system we can discern the entire system from knowing (at one time) the values of each of the $\mu_i$ and two of $A,B,\theta$, again with the third determined by the contact Hamiltonian as a constraint. The price we pay for this is that the $\mu_i$ are dynamical objects albeit with simple equations of motion, whereas the $C_i$ are constants. 

From equation (\ref{ContactGeneralA}) we can form a Lagrangian for our system in the same way as was done in the Kepler system. We note that the equations of motion for the dummy variables $z_i$ are given by the general form - see \ref{EL}. These are thus:
\be {z_i}' = -i V_i \mu_i^{-(i+1)} \ee
and thus the equations of motion for our system can be found from the extremization of $A_f$ subject to 
\be A'= \frac{\theta'^2}{2} - \frac{A^2}{2} - \sum_{i:V_i \neq 0} (i+1)\left(\frac{-{z_i}'}{i V_i} \right)^{\frac{i}{i+1}} V_i \ee
from which the Euler-Lagrange equations are (after some algebraic manipulation):
\ba \theta''&=&-A\theta' - \sum_{i:V_i \neq 0} \left( \frac{-{z_i}'}{i V_i} \right)^\frac{i}{i+1} \frac{\partial V_i}{\partial \theta} = -A\theta' -\sum_{i:V_i \neq 0} \mu_i^{-i}  \frac{\partial V_i}{\partial \theta} \theta' \nonumber \\
    {z_i}''&=& {z_i}'\left(A(i+1)+\frac{\partial V_i}{\partial \theta} \frac{\theta'}{V_i}\right) \rightarrow (\log z_i')' = (i+1)\left(A+(\log V_i)'\right)
\ea
and the contact Hamiltonian is that given in equation (\ref{ContactGeneral}). The equation match those derived from the contact Hamiltonian, and thus those of the equivalent system when written in the usual symplectic Hamiltonian or Lagrangian forms. 

\section{The N-Body System}
\label{Sect:Nbody}

Let us now consider the further generalization of the above to include more than two particles in our description, the so-called ``n-body problem" \cite{Albouy_2012,montgomery2014threebody,fejoz2021classical,Barbour2014,Barbour2015,jackman2015hyperbolic}. There is a long history of study of such problems, with particular attention paid to the case of gravitational attraction between the bodies. This is at the heart of modelling many interesting phenomena from the dynamics and stability \cite{2009Natur.459..817L,Murray:1999zd,Batygin:2008yv} of the solar system to the formation of galaxies \cite{Reif93thecomplexity,Bertschinger,Klypin_1999}.

For simplicity, we will assume that each of the particles has the same mass.  We will work in center of mass coordinates, so for $n$ particles in $\mathbb{R}^3$ we thus need $3n-3$ coordinates. We will transform from the usual Cartesian basis in which the positions are given as $x_1, ..., x_{3n-3}$ to a description in terms of the overall size of the system, $R^2=\sum_{l=1}^{3n-3} x_l^2$ and positions on a $3n-4$-sphere, $\theta_1,...,\theta_{3n-4}$. We refer to this sphere as `shape space' \cite{montgomery2014threebody,Barbour2014}. Thus the Lagrangian for this system can be written in terms of these variables, their velocities, and the metric on $S^{3n-4}$ induced by embedding the unit sphere in $\mathbb{R}^{3n-3}$, $g_{jk}$ and the potentials $V_i$ :
\be \L = \frac{\dot{R}^2}{2} + \sum_{j,k} \frac{R^2 g_{jk} \dot{\theta}_j \dot{\theta_k}}{2} - \sum_i C_i R^i V_i (\vec{\theta}) \ee
In this Lagrangian, the potentials will depend explicitly on the $\theta_j$ as the physical separations of the particles are functions of these. We note again at this point that a rescaling that fixes angles $\theta_i$ will affect all kinetic terms equally, thus we are in the same position as with the Kepler problem - the increase in the number of particles has no effect on the choice of scalings that leave the form of the Lagrangian unchanged. Thus we can treat these in the same manner as above. In the Hamiltonian formulation, following a Legendre transformation the Hamiltonian is written in terms of the conjugate momenta $P_R=\dot{R}, P_j = \sum_l R^2 g_{jl} \dot{\theta_l}$:
\be \H = \frac{P_R^2}{2} + \sum_{j,k} \frac{h_{jk} P_j P_k}{2R^2} +  \sum_i C_i R^i V_i (\vec{\theta}) \ee
where $h_{jk}$ is the inverse of $g_{jk}$, i.e. $\sum_l h_{jl} g_{lk} = \delta_{jk}$ where $\delta_{jk}$ is the Kronecker delta. Since rescaling affects all kinetic terms in the same way, we see that the dynamical similarities of this Hamiltonian are a further generalization of those given in equation (\ref{GenKeplerDS}):
\be \D_\zeta:= \sum_j P_{j} \frac{\partial}{\partial P_j} +\zeta R\frac{\partial}{\partial R} + (1-\zeta) P_R \frac{\partial}{\partial P_R} + \sum_{i:V_i \neq 0} (2-(2+i)\zeta) C_i \frac{\partial}{\partial C_i} \ee
As above we will work with the case where $\zeta=1$ for ease of comparison, though the construction is general, and absorb the energy $E$ into $V_0$, and thus work on the $H=0$ surface. We will express our system in terms of the invariants of $\D_1$:
\be A=P_R, \quad B_j = \frac{P_j}{R}, \quad \text{and for} \quad i \neq 0 \quad \mu_i = \frac{1}{C_i^{\frac{1}{i}}R}, \quad \mu_0 = C_0 \ee
In these terms the contact Hamiltonian and contact form are similar to those of equation(\ref{ContactGeneral}) :
\ba  \label{ContactNBody} 0=:&\H^c& = \frac{A^2}{2} + \sum_{j,k} \frac{h_{jk} B_j B_k}{2} + \mu_0 + \sum_{i:V_i \neq 0} \mu_i^{-i} V_i \\ &\eta& = -dA +\sum_j B_j d\theta_j + \sum_i \mu_i dz_i \ea 
The equations of motion we obtain from this are entirely analogous to those of equations (\ref{GenMuEom},\ref{GenBEom} and \ref{ContactGeneralA}) appropriately summed over indices. Of particular note is the equation of motion for $A$, which, after a Legendre transformation, forms the Herglotz-Lagrangian. In close correspondence with the above, this becomes:
\be \L^h = A' = \sum_{j,k} \frac{g_{jk} \theta'_j \theta'_k}{2} - \frac{A^2}{2} - \sum_{i:V_i \neq 0} (i+1)\left(\frac{-{z_i}'}{i V_i} \right)^{\frac{i}{i+1}} V_i \ee
where once again the $z'_i$ are velocities, related to the momenta $\mu_i$ through:
\be {z_i}' = -i V_i \mu_i^{-(i+1)} \ee
Thus we see a complete and closed description of the dynamics of the n-body problem can be written entirely in terms of the shape space, velocities thereon, and the variable $A$ which represents the apparent friction on shape space induced by changes of the overall size of the system in the Euclidean space. Such a description is important for several reasons; the first is that it may shed more light on the set of total collisions. These are characterised by $R\rightarrow 0$ in the decomposition we used for our Lagrangian. However, the Herglotz-Lagrangian (or contact Hamiltonain) description of this system makes no reference to scale. Therefore this description may be better suited to describing total collisions in terms of shape space, following e.g. \cite{Barbour2015}. A second point of interest is that there are points at which the system behaves as though there is no friction, those being the when $A=0$ which correspond to the overall size neither increasing nor decreasing. These are what Barbour refers to as ``Janus points" \cite{barbour2016janus,Gryb:2021qix}, though we note that Janus points are not universally points of zero expansion, but have a more complex role. They mark distinguished points of the evolution at which the system is instantaneously conservative. These are particularly interesting as places at which to evaluate the measures of solutions \cite{Corichi:2010zp,Corichi:2013kua,Measure,Measure2,Barbour2015} to assess whether a particular configuration can be considered `typical'. In our setup, as there is no reference to scale the configuration space is shape space and hence compact. This alleviates some of the problems of cut-offs and infinities that arise when attempting to measure typicality of such systems.

\section{Cosmology}
\label{Sect:Cosmology}

Cosmology offers a natural arena for examining the role of scale in physical systems. It is well-known that within the ubiquitous Friedmann-Lema\^itre-Robertson-Walker (FLRW) models the scale factor, $a$ must be fixed to some value at a given time, with the usual choice that the present value is set to unity. The choice of physical event at which to set this (or equivalently the value to which it is currently set) has no effect on physical observables. 

The FLRW models are symmetry reduced solutions to Einstein's equations. After enforcing homogeneity and isotropy on the spatial metric, the only remaining information is the behaviour of the scale factor, $a(t)$. The space-time metric is given:
\be ds^2 = -dt^2 + a(t) \left(dr^2 + f(r)(d\theta^2 + \sin^2 \theta d\phi^2)\right) \ee
where the function $f(r)$ depends on the curvature of the spatial slice; for flat spaces ($k=0$) $f(r)=r$, for closed spaces ($k>0$) $f(r)=\frac{\sin(\sqrt{k}r)}{\sqrt{k}}$ and for open spaces ($k<0$), $f(r)=\frac{\sinh(\sqrt{-k}r)}{\sqrt{-k}}$. 

We derive the dynamics for the system from the Einstein-Hilbert Lagrangian written in terms of the metric $g$ and Ricci scalar $R$ with matter determined by a Lagrangian $\L_m$;
\be S = \int \sqrt{g} (R+\L_m) d^4x \ee
wherein we have adopted the unit convention that $8\pi G=1$. Following the principle of symmetric criticality we can induce an action on the space of homogeneous, isotropic space-times. We choose a fiducial cell in space to integrate over (this is arbitrary since the spatial slices are homogeneous) and are left with
\be \label{CosmoAction} S = \int a^3 \left(-3\frac{\dot{a}^2}{a^2} +\frac{3k}{a^2} +\L_m(q,\dot{q}) \right) dt \ee
and from this the usual Euler-Lagrange equations give rise to the acceleration equation, and the Hamiltonian is the Friedmann equation, which we will express in terms of the Hubble parameter $H=\frac{\dot{a}}{a}$:
\ba \label{acceqn} 2\dot{H}+3H^2 + \frac{k}{a^2} = \L_m \\
 \label{Friedmann1}   H^2 + \frac{k}{a^2} = \frac{\H_m}{3} \ea
where $\H_m$ is the matter Hamiltonian obtained from the matter Lagrangian. Note that the inclusion of a cosmological constant can be achieved by adding a constant term to $\L_m$ (or equivalently $\H_m$).

 The freedom to fix the value of $a$ at any time is reflected in the rescaling of the action under: $a \rightarrow \lambda a, k \rightarrow \lambda^2 k$, in which case we find that $S \rightarrow \lambda^3 S$, for real positive $\lambda$, and the matter degrees of freedom remain unaffected. It is important to note that the Hubble parameter, $H$, is unaffected by this rescaling. In making this transformation we have again extended our dynamical similarity to act on the constant, $k$, and our equations of motion remain unchanged. Further we have kept the time coordinate, $t$ unchanged, thus we map between solutions with the same time parametrization. 

In previous work \cite{Sloan:2020taf}, we have treated the general case in which the matter Lagrangian, $\L_m$ is left as a general function of $q$ and $\dot{q}$. Here, for clarity of exposition and to allow more direct comparison to the usual cosmological literature, we will simplify the situation by  restrict ourselves to considering the matter to be a mixture perfect fluids with constant barotropic parameter. The matter Hamiltonian, $\H_m$ is usually decomposed into components, $\H_m= \sum_i \rho_i$, that have differing pressures, $P_i$, and thus differing dependence on the scale factor $a$. From the vanishing of the covariant derivative of the stress-energy tensor in general relativity, we arrive at the continuity equation
\be \label{Continuity} \dot{\rho_i} + 3H(\rho_i +P_i) = 0 \ee
For a perfect fluid the pressure is proportional to the energy density, $P_i = w_i \rho_i$, with $w_i$ the (constant) barotropic parameter. Hence solving equation (\ref{Continuity}) we see that:
\be \rho_i = \frac{\kappa_i}{a^{3+3w_i}} \ee
This allows us to do two things - the first is to treat the curvature term as a matter term for which $w=-1/3$. The second is to re-write equation (\ref{Friedmann1}) as the Friedmann equation in its more familiar cosmological form:
\be H^2 = \frac{8 \pi G}{3} \sum_i \frac{\kappa_i}{a^{3(1+w_i)}} \ee
wherein we have restored the Newton constant, $G$ for ease of comparison to the literature. This makes clear that we can re-write an action for such systems purely in terms of the scale factor and the constants:
\be \label{redact} S = \int a^3 \left(-3 \frac{\dot{a}^2}{a^2} + \sum_i \frac{k_i}{a^{3(1+w_i)}} \right) dt \ee
for some constants $k_i$. It is a simple exercise to show that the Euler-Lagrange equations for this system give rise to the usual Friedmann and acceleration equations. 

We are now in place to demonstrate the dynamical similarity within this system: as we noted above the action in invariant if we rescale both the scale factor, $a$, and the constants $k_i$ following
\be a\rightarrow \lambda a \quad k_i \rightarrow \lambda^{3(1+w_i)} k_i \quad \textrm{causes} \quad S \rightarrow \lambda^3 S \ee
Thus we can consider the quantities that are invariant under the transformation; $H=\frac{\dot{a}}{a}$ and $\frac{k_i}{a^{3(1+w_i)}}$. We will follow our method of promoting constants to dynamical variables to allow their description in a Herglotz action:  
\be \label{newredact} A' = \frac{3A^2}{2} + \sum_{w_i \neq 0} \dot{z}_i^{\frac{1+w_i}{w_i}} \ee
wherein $A=-H$. There are a few interesting points to note about this. The first is that this arises as an action principle which can be expressed as: `Extremize the Hubble parameter at time $t$, subject to the acceleration equation (\ref{acceqn}) and given an initial value $H_0$ at time $t_0$'. It should be no surprise that this is in close correspondence with the action described in \cite{Sloan:2020taf} as it was derived in the same manner, but for a simpler treatment of matter. The second is that the Herglotz-Lagrange equations for the $z_i$ become
\be \frac{d}{dt} \log(\dot{z_i}) = -3H w_i \ee
and hence, should we choose to reintroduce the scale factor $a$, then $\dot{z}_i \propto a^{-3w}$. However, this is not strictly necessary for the evolution of the system; the system itself can be completely integrated without ever referring to the scale factor, and can be shown to be integrable even in places where the symplectic system with scale factor is not \cite{Through,Sloan:2019wrz}. As we saw in the Kepler example, the scale itself is not a necessary quantity to include in our treatment. This can be carried forward by a Legendre transformation to the Hamiltonian description, where the contact Hamiltonian is 
\be \label{CosmoContact} \H^c = -\frac{3A^2}{2} + \sum_{w_i \neq 0} \frac{1}{w_i} \left(\frac{1+w_i}{w_i} \Pi_i \right)^{1+w_i} \ee
This makes clear why $w_i=0$ (the appropriate barotropic parameter for describing dust) has been excluded from our summation; its contribution to evolution arises as the value of the contact Hamiltonian, i.e. $\H^c = \rho_{\rm{dust}}$. Including this, we see that equation (\ref{CosmoContact}) is exactly the Friedmann equation (\ref{Friedmann1}) reproduced in these variables. 

In forming our dynamical similarity we made the choice to keep the time parameter fixed. This is not strictly necessary, as we could make the transformation:
\be a \rightarrow \lambda a \quad t \rightarrow \lambda^\mu t \quad k_i \rightarrow \lambda^{3(1+w)-2\mu} k_i \quad \textrm{causes} \quad S \rightarrow \lambda^{3-2\mu} S \ee
Following this process leads to an action that is equivalent to that of equation (\ref{newredact}) but written in a different lapse. In doing so, the value of the contact Hamiltonian will no longer correspond to the energy density of dust, but to some other matter component determined through the scaling. 

Finally, let us note the similarities and differences between the standard description of cosmology provided by the actions of equations (\ref{CosmoAction}) and (\ref{redact}). Dynamically, in terms of physical observables, where the symplectic system is well-defined the two descriptions are identical. An observer who measured, for example, the redshift of photons emitted by the CMB in either case would arrive at the same conclusions. In terms of physical ontology, and thus descriptive power, the two differ significantly. In the case of the former, we have to endow the universe with an unobservable quantity, the scale factor, and the differing fall-off of various matter types over time is the result of the change in this scale factor. In the latter case the differing behaviour is due to the frictional nature of the system; since it is an inherently dissipative system we should not be surprised that neither the total energy density, nor its components, are conserved. The same driving factor is at the mathematical root of both descriptions, it is the integral of the Hubble parameter over time. However in the original case this was used to describe the expansion of the universe whereas in the latter this is the amount of energy dissipated.  For each action the dynamics must be specified in terms of either the constants $\kappa_i$ at some time, or some values of the velocities $\dot{z}_i$ at some instant. Since the new formulation has a different set of basic elements, it is unlikely that quantizations of the two systems would remain identical - following Dirac's usual procedure of replacing Poisson brackets to commutators, for example, would differ significantly as the Hubble parameter has no conjugate variable in the new formulation.

In this work we have only considered a description of the FLRW cosmological models with simple matter sources. Both of these restrictions can be relaxed, and a more general prescription is given in \cite{Sloan:2020taf}, wherein the isotropy restriction is removed, considering homogeneous cosmological solutions wherein the spatial slice is a manifold of the type classified by Bianchi. Further, we leave the matter source as that which can be described through a general matter Lagrangian which is minimally coupled to gravity. The results therein are equivalent to that which we have described here, as the scaling of the action is closely related that which we have discussed. 

\section{Discussion}
\label{Sect:Discussion}

In this paper we have shown how scaling symmetries between theories can be expressed as dynamical similarities by extending our description to include the coupling constants of the theories as velocities (or momenta in the Hamiltonian framework). It is important to emphasise that this symmetry does not imply scale invariance of physical observables, but that there is a redundancy in the mathematical description of our theories which we can eliminate. We have shown that the elimination of this redundancy reveals that these systems can be described in terms of a frictional system, whose dynamics can be derived following an action principle of the Herglotz type. 

The idea of rescaling the constants of our theory may seem somewhat esoteric. One could argue that since we observe specific values of, for example, Newton's constant, in our universe we should restrict ourselves to descriptions of reality which use only this value. However, such arguments are based upon a false premise - that it is possible to uniquely determine these constants from within a system determined by their values. Rather, these constants are determined by making observations of the universe itself. Hence in a strict sense they are relationally determined. Newton's constant can be determined following a torsion balance experiment designed by Cavendish \cite{Cavendish}. This is a physical device whose dynamics are determined by similar laws to those described in equation (\ref{HookeNewton}), though the geometric set-up is notably more complicated. Nonetheless, the same scaling can be applied to reveal indistinguishable observations. Therefore in considering the space of physical theories which cannot be distinguished from one-another by observation, we are justified in considering transformation of couplings alongside changes of positions and velocities. 

We have considered three physical systems as exemplars of scaling symmetries: The Kepler problem (and its generalizations), the n-body problem, and homogeneous, isotropic cosmologies. In each case we see that there is a frictional description of the dynamics which makes no reference to the original scale. This description is autonomous; the equations of motion of the scale invariant quantities are do not depend on scale, and can be derived from an action principle which also makes no reference to scale. Such descriptions are particularly interesting at points where the scale of the system proves problematic in the usual framework such as the total collision of the n-body system, and the initial singularity of cosmology. In such cases the evolution of the system can become ill-determined as, for example, the equations of motion are no-longer Lipschitz continuous, and hence the uniqueness of solutions cannot be guaranteed. In the case of cosmology, it has been shown that the reduced, scale-free description does not suffer from such problems, and thus can be continued beyond these points \cite{Through,Sloan:2019wrz}. This may be a hint that the scale-free description is more fundamental, and hence potentially a more fertile starting point for quantum theories of gravity. 

The frictional nature of our descriptions is seen through the focussing of measures, following equation \ref{MeasureDissipation}. This has important implications for discussions of the arrow of time (see e.g. \cite{Gryb:2020wat,Barbour2014,Barbour2015}). In  several systems the focussing is monotonic for large periods, such as flat or open cosmologies and the n-body problem with gravitational forces and positive total energy. This can be used to explain why there is an apparent thermodynamic arrow of time, which points in the direction of measure focussing. Since we can always `symplectify' a contact system \cite{ArnoldBook}, by introducing scale into our description we arrive at a conservative system. In such cases, Liouville's theorem applies, and hence measures on phase space like the natural measure $\Omega$ of section \ref{Sect:LagMech} are preserved. This preservation is brought about precisely by expanding or shrinking the extent of the measure along the unobservable scale direction to compensate for the focussing on the observable directions. 

\section*{Acknowledgements}

The author is grateful to Alessandro Bravetti, Sean Gryb and Flavio Mercati for their comments on a draft of this paper.

\appendix

\section{Derivations of the Kepler System}
\label{Appendix}

In this appendix, we will present a complete derivation of the systems described in section \ref{Sect:KeplerExample}. In doing so we will not rely directly upon the well-known Euler-Lagrange equations, and their less well-known contact counterparts, but instead show the explicit extremization of actions and derivation of Hamiltonian vector fields etc. Our reason for this is two-fold. In the first instance this will provide a clear example of the similarities and differences between the two approaches and how they manifest in a physical system. Further this will show the workings of a contact system in a more easily context than the abstract use of, e.g. Darboux coordinates. 

We first begin with the Lagrangian of equation (\ref{KeplerLagrangian}). At each point on the tangent bundle, we consider the transformations
\be 
\Delta: \{r,\dot{r},\theta,\dot{\theta}\}\rightarrow \{r+\delta r, \dot{r}+\delta \dot{r}, \theta + \delta \theta, \dot{\theta}+\delta \dot{\theta}\}
\ee
The requirement that $\dot{x}$ represents a time derivative of $x$ forces $\delta \dot{x} = \frac{d}{dt} (\delta x)$ for configuration variables $x$. Minimizing the action under such transformations we see (dropping boundary terms which are specified by initial conditions and therefore cannot vary):
\ba 
0&=&\delta S = \int_{t_i}^{t_f} \left( \dot{r}\delta\dot{r}+(2r\dot{\theta}^2-\frac{C}{r^2})\delta r + r^2 \dot{\theta} \delta \dot{\theta} \right)dt \nonumber \\
&=& \int_{t_i}^{t_f} \left( (-\ddot{r} + r^2 \dot{\theta}^2 -\frac{C}{r^2})\delta r - \frac{d}{dt} (r^2 \dot{\theta}) \delta \theta  \right) dt 
\ea
Setting the coefficients of $\delta r$ and $\delta \theta$ to zero gives rise to the usual equations of motion of the Kepler system. 

Let us now compare with the system described through Herglotz principle: The first significant change is that we only have three variables, $A,\theta$ and $\theta'$. In the same way that we had to relate the variations in position and velocity in the usual Lagrangian system, we note that $\delta \theta' = \frac{d}{d\tau} \delta \theta$. Further to ensure that we retain the relationship between $A$ and $A'$ we note that at each instant in the evolution $\delta A = \frac{\partial A'}{\partial \theta'} \delta \theta$. For a rigorous derivation of this, see e.g. \cite{ArnoldBook,deleon2021herglotz}. Once again we ignore boundary terms to arrive at 
\ba
\nonumber 0 &=& \delta A_f = \int_{t_i}^{t_f} \left( \theta' \delta \theta' - A \delta A \right) dt \\
&=& -\int_{t_i}^{t_f} \left(\theta'' + A \theta' \right)\delta \theta \, dt
\ea
Hence minimization of this action gives exactly the dynamics of our system.

Let us now compare the case of the Hamiltonian systems. In regular usage we would express our system in Darboux coordinates and use the well-known Hamilton equations to find the evolution of the system, or find the evolution of variables by taking Poisson brackets with the Hamiltonian. However we are again going to derive this in terms of vector fields on a phase-space manifold. We show this so that the inner working of both the Hamiltonian phase space description and the contact Hamiltonian system can be compared. 

The Hamiltonian vector field $\X_\H$ satisfies \cite{ArnoldBook}:
\be \label{HamiltonianEOMDefn} \iota_{\X_\H} \omega = -d\H \ee
and hence we can read off the equation of motion for each of our phase space variables by comparing these one-forms. If we express the coefficients of the vector field in terms of the phase space variables we can write
\be \X_\H = \X_{P_r} \frac{\partial}{\partial P_r} + \X_{r} \frac{\partial}{\partial r} +  \X_{P_\theta} \frac{\partial}{\partial P_\theta} +  \X_{\theta} \frac{\partial}{\partial \theta}  \ee
Thus we find from equation (\ref{SymplecticKepler}) the left hand side of equation (\ref{HamiltonianEOMDefn}) is:
\be \iota_{\X_\H} \omega = \X_{P_r} dr - \X_r dP_r + \X_{P_\theta} d\theta - \X_\theta dP_\theta \ee
and the right hand side is:
\be d\H = P_r dP_r + \frac{P_\theta}{r^2} dP_\theta + \left(\frac{C}{r^2} - \frac{P_\theta}{r^3} \right)dr \ee
and hence comparing coefficient of the base one-forms we see:
\be \dot{r}=P_r \quad \dot{P_r}=\frac{P_\theta^2}{r^2}-\frac{C}{r^2} \quad \dot{\theta}=\frac{P_\theta}{r^2} \quad \dot{P_\theta}=0 \ee
Let us now contrast this with the derivation of the equations of motion from the contact Hamiltonian system given in equation (\ref{ContactKepler}). Since our system is already expressed in Darboux coordinates, we could rely on the standard contact equations of motion given in equation (\ref{CHEoM}). However in more complex systems it may be preferable to use dynamical similarity to eliminate, for example, the dilation of objects used to define a rod, which in turn will not necessarily render the system in Darboux coordinates. Hence below we demonstrate in more detail how the dynamics of the system can be calculated as a flow on the extended contact phase-space.

Given a one-form $\eta$ there is a vector $\R$, called the Reeb vector, on the contact space satisfying:
\be \label{Reebdef} \eta(\R) =1 \quad \iota_\R d\eta = 0 \ee
and in Darboux coordinates this is given by $\R=\pv{S}$.  The contact Hamiltonian vector field $\Y$ satisfies \cite{Bravetti2}
\be \label{ContactEOMDefn} \iota_\Y d\eta +(\iota_\Y \eta) \eta = -d\H^c + (\iota_R d\H^c +\H^c)\eta \ee
In our set up, the contact Hamiltonian is a constraint, with the system arranged such that $\H^c=0$ throughout. To find the evolution of the system, we consider a general vector field on the extended contact space:
\be \Y= \Y_A \pv{A} + \Y_B \pv{B} + \Y_\theta \pv{\theta} + \Y_\mu \pv{\mu} +\Y_z \pv{z} \ee
The left hand side of equation (\ref{ContactEOMDefn}) is then:
\be
\iota_\Y d\eta +(\iota_\Y \eta) \eta =-(\iota_\Y \eta)dA -\Y_\theta dB + (\Y_B + B \iota_\Y \eta) d\theta - \Y_z d\mu +(\Y_\mu +\mu \iota_\Y \eta)dz 
\ee
and the right hand side is:
\be 
d\H^c - (\iota_R\H^c +\H^c)\eta = BdB+ABd\theta -d\mu +A\mu dz 
\ee
Hence equating coefficients and subtracting the Hamiltonian constraint (which is zero) to $A'$ for clarity, we see:
\be A'=\frac{B^2}{2}-\frac{A^2}{2}+1 \quad B'=-AB \quad \theta'=B \quad \mu'=-A\mu \quad z'=1 \ee

\bibliographystyle{apsrev4-2}
\bibliography{HerglotzCouplings}

\end{document}